\documentclass[fleqn,usenatbib,usedcolumn]{mnras}
\usepackage{amsmath}
\usepackage[british]{babel}
\usepackage{txfonts}

\volume{{\rm in press}}
\usepackage[T1]{fontenc}
\usepackage{ae,aecompl}
\usepackage{multirow}
\usepackage{array}
\usepackage{enumerate}
\usepackage{appendix}
\usepackage{graphicx}
\usepackage{graphics}
\usepackage{epstopdf}

\newcommand{\yr}{\mbox{ yr}}

\newcommand{\Gyr}{\mbox{ Gyr}}

\newcommand{\pc}{\mbox{ pc}}

\newcommand{\Msun}{M_{\odot}}

\newcommand{\sref}{\S~\ref}

\newcommand{\comment}[1]{}

\title[The Way To a Double Degenerate]{The Way To a Double Degenerate: $\sim15-20$ per cent of $1\Msun \le M \le 8\Msun$ Stars have a $M>1\Msun$ Companion}

\author[Klein \& Katz]{
	Ygal Y. Klein,$^{1}$\thanks{E-mail: ygalklein@gmail.com}
	Boaz Katz$^{1}$
\\
	$^{1}$Dept. of Particle Phys. \& Astrophys., Weizmann Institute of Science, Rehovot 76100, Israel\\
}

\pubyear{2016}

\date{Accepted XXX. Received YYY; in original form ZZZ}

\begin{document}
\label{firstpage}
\pagerange{\pageref{firstpage}--\pageref{lastpage}}
\maketitle

\begin{abstract}
	We find that $\sim 15-20$ per cent of A-type stars or red giants are bound with a massive companion ($M_{\rm secondary} > 1\Msun$) in an intermediate wide orbit ($0.5<P<5000\yr$). These massive binaries are expected to form wide-orbit, double-degenerate systems (WODDs) within $\lesssim10\Gyr$ implying that $\sim10$ per cent of white dwarfs (WDs) are expected to be part of a WODD with a lighter WD companion. These findings are based on an analysis of previous adaptive optics observations of A-type stars and radial velocity measurements of red giants and shed light on the claimed discrepancy between the seemingly high multiplicity function of stars and the rather low number of detected double degenerates. We expect that GAIA will find $\sim 10$ new WODDs within $20\pc$ from the sun. These results put a stringent constraint on the collision model of type Ia supernovae in which triple stellar systems that include a WODD as the inner binary are required to be abundant.
\end{abstract}

\begin{keywords}
	White Dwarfs -- Binaries: general -- Supernovae: Type Ia
\end{keywords}

\section{Introduction}\label{sec:Introduction}

Type Ia Supernovae (SNe) are among the most luminous and energetic events observed. Following decades of extensive observational surveys and modeling efforts, there is good evidence that these events are the result of thermonuclear explosions of carbon oxygen white dwarfs (CO-WDs) but it is still unknown what triggers $\sim 1$ per cent of them to explode \citep[for a recent review, see e.g.][]{maoz14}. 

One of the scenarios recently argued to be the progenitor of type Ia SNe is the direct collision (as opposed to merger) of two CO-WDs \citep{katz12, kushnir13, dong15}. Following previous demonstrations that colliding WDs explode \citep{rosswog09, raskin10, hawley12}, \cite{kushnir13} showed numerically that such collisions with the observed range of WDs masses robustly lead to thermonuclear explosions with the observed range of brightness and late time characteristics. \cite{dong15} reported observations of double peaks in the spectra of some events, a unique prediction (so far) of the collision model.

Until recently, the rate of direct collisions was considered to be orders of magnitudes lower than the type Ia rate \cite[e.g.][]{rosswog09, raskin09}. \cite{thompson11} recently argued that the merger rate of WDs due to gravitational waves may be enhanced in triple systems by the Lidov-Kozai mechanism and noted that some direct collisions may also occur in such systems. It was later shown by \cite{katz12} that the rate of WD direct collisions may be as high as the type Ia rate if tens of percents of WDs reside in (mildly) heirarchical triple systems with a wide-orbit-double-degenerate (WODD hereafter) inner binary (semimajor axis $1\lesssim a_{\rm in} \lesssim 1000$AU), raising the possibility that most type Ia SNe are due to direct collisions.

A critical requirement for the collision model is that a sufficient amount of triple systems with the required hierarchy exists. In particular, such systems should have an inner WODD. A first step to determine the abundance of such relevant triple systems is to find out the abundance of WODDs. In this paper we therefore attempt to answer the following question: \textbf{what is the fraction of CO-WDs that have a lighter CO-WD companion with a wide orbit ($\mathbf{P\gtrsim1\yr}$)?}.

A straight forward approach to answer this question is to examine the population of WDs in the solar neighborhood. This approach was presented in \cite{holberg09} based on the local sample of WDs within $D<20\pc$ claimed to be $80$ per cent complete by the authors \citep{holberg08}\footnote{The sample has been updated to $136$ WDs without any new WODDs and its current completeness estimate by the authors is $86$ per cent \citep{holberg16}.}. In this sample, there are only $3$ WODDs\footnote{WD-0121-429 is an additional uncertain candidate. One should note that the separation of WD-2226-754 is slightly larger $\sim 1400$AU \citep{scholtz02} but is still counted as a WODD. There are $3$ additional double degenerate systems in the local sample \citep{holberg08}, however WD-0322-019 (G77-50) was found to be a single star \citep{farihi11} and two other systems are short period close binaries ($P\sim2$ days). WD-0532+414 which is newly entering the local ($D<20\pc$) sample in the current version is a short period close binary based on its radial velocity (RV) measurements \citep{zuckerman03}.} out of $136$ WDs in total. This count suggests that only $\mathbf{\sim 2}$ per cent of WDs have a (lighter, wide orbit) WD companion (and only $\sim 30$ per cent have any companion \citep{holberg16}).

This result is very low compared to the binarity fraction ($\sim70-100$ per cent) of the progenitors of todays WDs - intermediate mass main sequence (MS) stars \citep[M $\sim1.5-8 \Msun$, e.g.][]{kobulnicky07, kouwenhoven07} and their supposed mass ratio distribution of $f(q)\propto q^{-0.5}$. Moreover, $4$ out of the closest $6$ WDs are in binary systems \citep{holberg16} and the two closest WDs - Sirius B and Procyon B have massive ($M\gtrsim 1.5 \Msun$) MS companions that will become WDs within $\sim 1\Gyr$ and are thus likely to become WODDs \citep{liebert05, liebert13}. If the fraction of WODDs is indeed $\sim2$ per cent, this is a strange (but possible) coincidence. Another option is that for some reason, many of the wide-orbit MS massive binaries do not end up as WODDs. Interaction during the stellar evolution is unlikely to play a significant role beyond separations of a few AU and we assume that bound systems remain bound (however see comment about this assumption in section \sref{sec:Discussion}). These puzzles have led to suggestions that \cite{holberg08} is not as complete as reported by the authors \citep{ferrario12, katz14}.

In this paper we attempt to quantify the expected fraction of WODDs based on observations of the relevant WD progenitors - intermediate mass ($1<M<8\Msun$) MS stars that will become WDs within a Hubble time. An adaptive optics (AO) survey of A-type stars within $75\pc$ was recently conducted by \cite{DeRosa14} allowing the binarity fraction at long periods ($P\gtrsim50\yr$) to be established. In particular the relevant massive ($M_{\rm secondary}>1\Msun$) companions have sufficiently low contrast to be reliably detected. This is discussed in section \sref{sec:VAST_AO}. The fraction of companions with shorter periods is more challenging. Finding binaries with periods $P\sim 1-10\yr$ is currently best achieved by radial velocity (RV) surveys. However, the rapid rotation of the relevant intermediate mass stars broadens the lines and makes it very challenging \citep{verschueren99}. A way around this problem is to observe these stars when they are in the red giant phase in which the rotational broadening is greatly reduced. An extensive RV survey of red giant stars in open clusters was preformed by \cite{mermilliod08} providing an excellent sample. Again, the fact that only companions with significant mass are considered implies large signals increasing our confidence of detection. An analysis of the sample is done in section \sref{sec:RVs}.

\emph{We find that $\mathbf{\sim15-20}$ per cent of massive stars have a (lighter) massive companion $M_{\rm secondary} > 1\Msun$ in the period range $0.5\lesssim P\lesssim 5000\yr$ with a uniform distribution in logarithmic space or equivalently $\mathbf{\sim4}$ per cent per dex in period (see Fig.~\ref{fig:massive_binaries_occurence_vs_period_compilation_M2_gt_1})}. A roughly flat distribution in log space is typical for wide binaries \citep[e.g.][]{raghavan10} and the fact that such a distribution is obtained increases our confidence in the estimate which is based on very different samples at the two ends of the period range which covers $4$ orders of magnitude.

\section{Adaptive Optics measurements of A-type stars}\label{sec:VAST_AO}

\begin{figure}
	\includegraphics[scale=0.5]{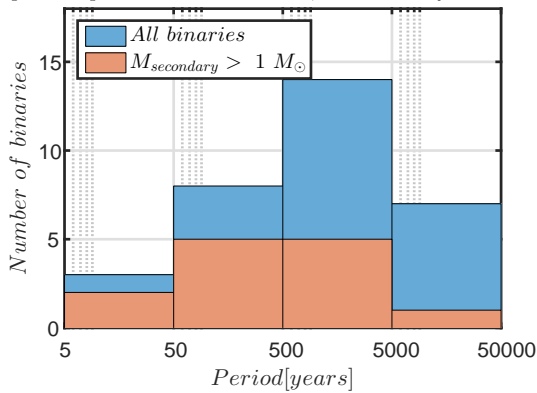}
	\caption{\label{fig:period_histogram_A_stars_D_lt_50_D_gt_30_M2_gt_1}
		Binary companions to A-type stars in the distance range $30<D<50\pc$ observed by \protect\cite{DeRosa14} using AO. The period is obtained from the projected separation, $a_{\rm proj}$, using $P\equiv\left(\frac{(a_{\rm proj}/AU)^3}{(M_1 + M_2)/\Msun}\right)^{1/2} \yr$. The brown shading is the number of binaries in the period bin with a massive ($M_{\rm secondary}>1\Msun$) secondary. 
	}
\end{figure}

\cite{DeRosa14} conducted an adaptive optics (AO) survey of $363$ A-type stars (identified by photometric colour and brightness by \cite{hog00}) drawn from a volume limited sample selected from the \textit{Hipparcos} catalogue \citep{ESA97, vanleeuwen07} within $75 \pc$ from the sun. The distance distribution of the Hipparcos sample indicates that it is complete within $D\lesssim50\pc$ \citep{DeRosa14}. The massive companions with $M_{\rm secondary}>1\Msun$, have a contrast in \textit{K}-band of $\Delta m<3$ mag compared to the A-type stars primaries. Therefore, according to fig. 8 of \cite{DeRosa14} such massive binary systems are detectable with confidence $\gtrsim80$ per cent for an angular separation range of $0.3\lesssim \rho \lesssim15$ arcsecs, corresponding to projected separations of $15\lesssim a_{\rm proj}\lesssim 750$ and $9\lesssim a_{\rm proj} \lesssim 450$AU at $D=50$ and $30\pc$ respectively. We henceforth consider only systems with $30<D<50\pc$, obtaining a rather complete survey for a conservative projected separation range of $20\lesssim a_{\rm proj} \lesssim 420$AU corresponding to a period range of $50<P<5000\yr$\footnote{Adopting a conversion from projected separation to orbital period $P\equiv\left(\frac{(a_{\rm proj}/AU)^3}{(M_1 + M_2)/\Msun}\right)^{1/2} \yr$ and using a typical total mass of $M_1+M_2=3\Msun$ for a massive binary with an A-star primary one gets $P=50 \yr(a_{\rm proj}/20 AU)^{3/2}$. The precise conversion to orbital period is not important given the wide distribution in separations.}. There are $179$ A-type stars in the Hipparcos catalogue in this range and our sample consists of the $121$ among these that were observed by AO. The period histogram of companions to the A-type stars observed by the AO survey within $30<D<50\pc$ is shown in Fig.~\ref{fig:period_histogram_A_stars_D_lt_50_D_gt_30_M2_gt_1} using masses estimated by \cite{DeRosa14}. \emph{We find that $\sim \mathbf{4}$ per cent of A-type stars have companions with $M_{\rm secondary}>1\Msun$ in each of the logarithmic orbital period bins $50<P<500$ and $500<P<5000$}$\yr$. For the period bin of $5<P<50\yr$ we find a lower limit of massive binarity fraction of $\sim 2$ per cent. While we are interested in separations $a\lesssim 1000$AU, we note that in addition to the AO survey, \cite{DeRosa14} reported common proper motion (CPM) binaries at larger separations $a\gtrsim 3000$AU ($P\gtrsim 10^{5}\yr$) and found only one companion with $M_{\rm secondary}>1\Msun$ in this range. The total binarity fraction (including all companions) is decreasing for such wide separations as shown there (fig. 9).

\section{Radial Velocity Measurements of Red Giants}\label{sec:RVs}

Binaries with low projected separations $\rho\lesssim 0.1$ arcsecs are not resolved by the AO survey. Accurate radial velocity measurements of intermediate mass MS stars $2 \lesssim M \lesssim 8\Msun$ is challenging to obtain due to their typical fast rotation \citep{verschueren99}. This problem can be bypassed by observing the stars during their red giant phase where accurate RV measurements can be obtained. Red giants in open clusters are particularly useful given their known distance and the fact that their mass can be inferred from the age of the cluster.

An extensive RV survey of red giants in open clusters was concluded in \cite{mermilliod08} who obtained $10517$ measurements of $1309$ red giants over $\sim 15$ years. We limit our analysis to the sample of red giants marked as members of open clusters (with a known age) by \cite{mermilliod08} and measured more than once, resulting with $797$ giants. Orbital solutions for binary systems with $P\lesssim 15\yr$ were obtained and presented in \cite{mermilliod07}. The amount of companions in the period range $0.5<P<5\yr$ is shown in Fig.~\ref{fig:mass_histogram_red_giants_M2_gt_1} as a function of the primary mass. The primary mass is estimated as
\begin{equation}
M_{\rm RedGiant}= 10^{1/3}\Msun(\rm Age/\Gyr)^{-1/3}
\label{eq:M_t}
\end{equation}
\footnote{Only an approximated age-mass relation is needed given the weak dependence on primary mass. Eq. \eqref{eq:M_t} is obtained by assuming that $10$ per cent of the total mass is burned during the MS phase at a constant luminosity given by $L=\frac{7 \times 10^{-3} \times 0.1 \times Mc^{2}}{Age}$, consistent with the observed mass-luminosity relation \cite[e.g.][]{torres10}. The estimate in equation~(\ref{eq:M_t}) agrees to an accuracy better than $6$ per cent with the five red giants with mass evaluations given by \cite{mermilliod07}.} where the ages of the clusters are adopted from \cite{Dias14}\footnote{Except for the following clusters: IC-1396 from \cite{kharchenko05}, Collinder-258 is named Harvard-5, NGC-3247 from \cite{ahumada03} and NZ-Zor-1's mass was taken from NGC-6067 age according to \cite{majaess13}.}. Given that the primary's mass is known but the inclination of the orbit is not, the RV observations provide a lower limit for the companion's mass. The numbers of binaries with a minimal companion's mass above $1\Msun$ are shown in the figure. As shown in the figure, $29$ out of the $797$ red giants observed, {$\sim 3.6$ per cent, are bound in an orbit with period $0.5<P<5\yr$ and a companion with a minimal mass of $1\Msun$.
	
\begin{figure}
	\includegraphics[scale=0.35]{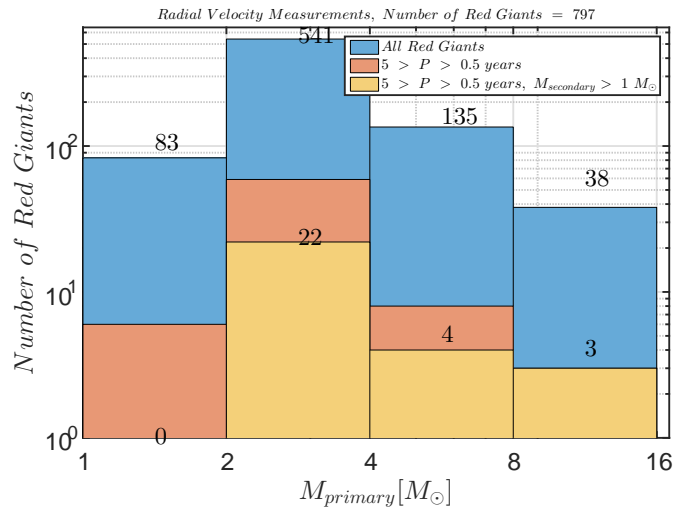}
	\caption{\label{fig:mass_histogram_red_giants_M2_gt_1}
		Binary companions to red giants in open clusters found by the RV survey of \protect\cite{mermilliod07}. The blue shading (and number written above) is the total number of red giants in each mass bin. The brown shading is the number of red giants within each mass bin for which a companion with an orbit in the range $0.5<P<5\yr$ was found and the yellow shading shows the number of such systems for which the minimal mass of the companion is $>1 \Msun$. 
		}
\end{figure}
	
Massive companions can be missed in this survey for two main reasons: 1. The inclination is too high so that while the companion's mass was above the threshold of $1\Msun$, the derived minimal mass was not. 2. The RV signal's amplitude was not high enough to allow for a detection and accurate solution and the system was misclassified. As we next show, the real fraction of massive companions is likely $\sim 1.3-2.1$ times larger than the observed fraction of systems with $M_{\rm minimal}>1\Msun$ implying a fraction of $\mathbf{\sim5-8}$ per cent.
	
The probability that the derived minimal mass for the secondary is lower than $1\Msun$ is plotted in Fig.~\ref{fig:p_for_missing_massive_m2_mermilliod_articles_lower_limit_1} as a function of the secondary's ('real') mass, assuming an isotropic orientations distribution (uniform distribution in $-1\le \cos i \le1$) for several typical values of the primary's mass. As can be seen in the figure, the probability that a companion with a mass of $M_{\rm secondary} = 2\Msun$ has an inclination that leads to a derived minimal mass $<1\Msun$ is $<0.2$. Naturally, the lower the companion's mass (above $1\Msun$) the higher the chance to miss it due to high inclination. The fraction of massive binaries that result with a derived minimal mass $<1\Msun$ due to inclination can be estimated given an assumed mass ratio distribution. We consider two mass ratio distributions: a. Uniform - $dN/dq=\rm const.$ \citep[e.g.][]{raghavan10} b. 'Kroupa IMF'- $dN/dq\propto q^{-2.3}$, which is obtained if the mass of the companion is drawn from the Initial mass function (IMF) suggested by \cite{kroupa01}. Assuming a typical primary mass of $\sim 3\Msun$ (the median mass of the red giants in the sample), we find that $\sim 25$ per cent (for the uniform distribution) and $\sim 37$ per cent (for the Kroupa IMF distribution) of massive binaries $M_{\rm secondary}>1\Msun$ would have a derived minimal mass below $1\Msun$ due to inclination. We therefore estimate that the real fraction of massive binaries for the period bin $0.5<P<5\yr$ is $\sim 1.3-1.6$ larger than the observed definite fraction obtained by our conservative restriction to a minimal mass larger than $1\Msun$ (due to unknown inclination).
	
\begin{figure}
	\includegraphics[scale=0.5]{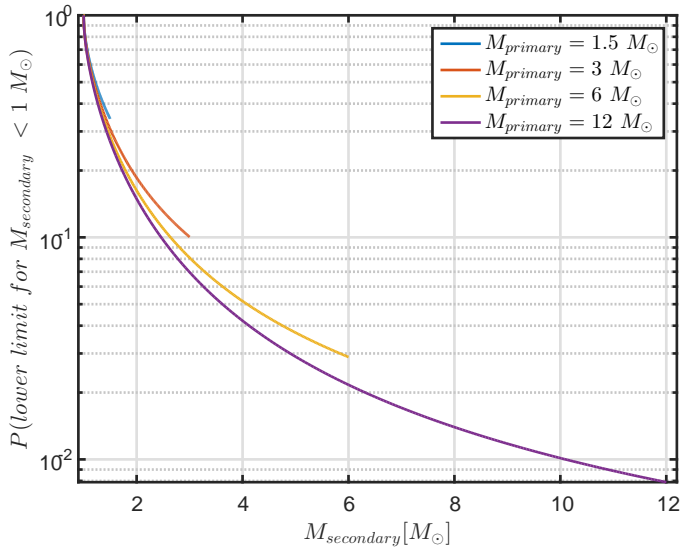}
	\caption{\label{fig:p_for_missing_massive_m2_mermilliod_articles_lower_limit_1}
		The probability that a binary system with a primary $M_{\rm primary}$ and a secondary $M_{\rm secondary}$ is sufficiently inclined so that the minimal mass deduced from an RV measurement of the primary is $<1\Msun$. Different curves show the probability as a function of the secondary mass for different primary masses.
		}
\end{figure}
	
In order to estimate the amount of massive binaries that may have been missed due to a low signal we preformed the following conservative Monte Carlo analysis. We examined the peak-to-valley (PTV) RV variations. We divide the whole sample into two distinct populations - systems with observed low PTV $< 10\rm km s^{-1}$ and high PTV $> 10\rm km s^{-1}$. In the high PTV group, where the vast majority of systems are solved, we have $41$ definite massive binaries ($29$ of them in our period range), $46$ systems with a solved orbit but a derived minimal mass $<1\Msun$ and $23$ unsolved systems. We reviewed the data for each of these $23$ candidates and found that $13$ of them are potential genuine binaries with too little observations to determine the minimal mass and period\footnote{The ten systems excluded are: NGC-663-319, NGC-2099-12, NGC-2099-255, NGC-6940-188, NGC-2324-2555, NGC-6067-240, NGC-6208-48, NGC-6664-53, NGC-2099-92 and Melotte-71-29.}. Given that about a third of the solved binaries with high PTV have $M_{\rm minimal}>1\Msun$ and $0.5<P<5\yr$, we expect $\sim 5$ of these systems to be relevant massive binaries. For each of the systems with PTV$<10\rm km s^{-1}$, we estimated the probability $P_{\rm PTV}$ that a RV curve with a minimal mass of $1\Msun$ would result with a PTV value which is lower than the one observed due to the limited sampling of the curve. This was done by calculating for each such system, $10^4$ RV synthetic signals at the observed phases by assuming the existence of a massive companion $M_{\rm secondary}=1\Msun$ on an edge-on orbit ($\sin i =1$) with a period drawn from a uniform distribution in log space in the range $0.5<P<5 \yr$, eccentricity drawn from a uniform distribution (throwing out systems with periastron $<0.5$AU), uniform orientation $0\le \omega<2\pi$ and initial phase $0\le T < P$. The probability $P_{\rm PTV}$ has a uniform distribution and an expectation value of $<P_{\rm PTV}>=0.5$. In practice, the expectation value is larger due to measurement errors (which tend to increase the PTV). An upper limit for the number of missed systems is thus obtained by $N=2\times \sum_{\rm systems} P_{\rm PTV,i} \sim 17$. Assuming $\sim 15$ missing systems due to inclination, $\sim 5$ missing systems in the high PTV group and $\lesssim 15$ missing systems in the low PTV group, we obtain an upper limit of $\sim 65$ systems in total or $\sim 8$ per cent of the sample. Assuming only the expected $\sim 20$ missing systems due to unknown inclination, implying $\sim 50$ systems with massive companions in the period range $0.5<P<5 \yr$, allows a $1 \sigma$ lower limit due to Poisson statistics of about $(50-\sqrt{50})/797\approx 0.05$. The observed fraction of systems $29/797 \sim 0.035$ is thus a conservative lower limit.

\section{Discussion}\label{sec:Discussion}

In this paper we analysed previous AO observations of A-type stars by \cite{DeRosa14} (section \sref{sec:VAST_AO}) and RV measurements of red-giants in open clusters by \cite{mermilliod08} (section \sref{sec:RVs}) to obtain a robust estimate of the fraction of massive stars $1<M<8 \Msun$ that have (lighter) $M > 1\Msun$ companions in a wide orbit ($P\gtrsim 1\yr$). Assuming that these systems will remain in-tact when the stars evolve to become WDs within $\lesssim 10\Gyr$, they will become wide orbit double degenerate systems (WODDs). The results for the two populations are shown in Fig.~\ref{fig:massive_binaries_occurence_vs_period_compilation_M2_gt_1}. As can be seen, the binaritiy fraction per logarithmic period bin is rather constant across $4$ orders of magnitude $0.5<P<5000\yr$ using different techniques. About $\sim15-20$ per cent of massive stars have such massive companions in this period range with about $\sim4$ per cent for each dex of period. The samples are likely close to being complete in this range given the high luminosity (for large separations) and large RV signal (for low separations) as demonstrated in sections \sref{sec:VAST_AO} and \sref{sec:RVs} (except for the $5<P<50\yr$ bin where we obtain a lower limit for the fraction $\sim 2$ per cent). The upcoming release by GAIA, expected in September 2016, should confirm the results based on the AO observations at large separations with much larger statistics. In particular, by providing parallax and proper motion for the Tycho 2 catalogue, main sequence (MS) stars with $M>1\Msun$ should be measured to over $100\pc$.

\begin{figure}
	\includegraphics[scale=0.37]{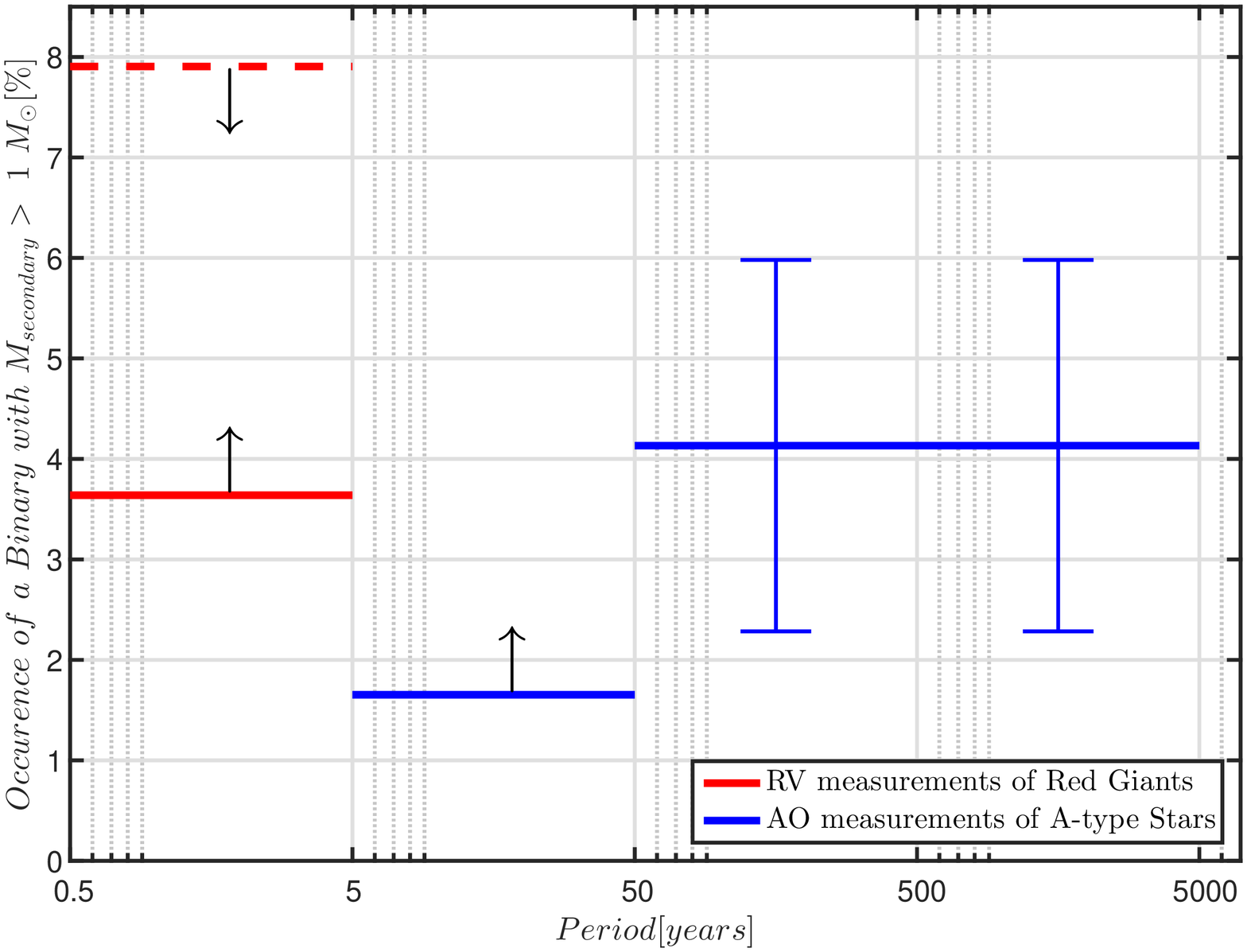}
	\caption{\label{fig:massive_binaries_occurence_vs_period_compilation_M2_gt_1}
		The fraction of massive stars that have lighter companions with $M_{\rm secondary}>1\Msun$ as a function of the orbital period. The red solid line and red dashed line represent lower and upper bounds respectively from RV measurements of red giants in open clusters in the logarithmic period bin $0.5<P<5\yr$. Blue solid lines with error bars (1$\sigma$ statistical) represent the fraction obtained from AO measurements of A-type stars in two logarithmic bins in the range $50<P<5000\yr$ (based on 5 detected systems in each bin). The blue line in the period bin $5<P<50\yr$ is a rough lower limit (based on the 2 systems detected) in this range which is only partly covered by the AO survey.
	}
\end{figure}

Based on these results we expect that \emph{$\mathbf{\sim15-20}$ per cent of WDs have wide orbit ($P\gtrsim 1\yr$) companions which are either (lighter) WDs or massive MS stars ($M>1\Msun$)}\footnote{This does not necessarily apply to low mass WDs, $M_{\rm WD}\lesssim 0.55 \Msun$ whose progenitors with mass $M \lesssim 1.5 \Msun$ were not probed by the samples presented here.}.

In order to estimate the fraction of WDs that have a wide orbit WD companion, the fraction of stars with $M>1\Msun$ that have already evolved into WDs needs to be estimated. Assuming the age of our galactic disk is $\sim 9.5 \Gyr$ \citep{oswalt96}, a constant star formation rate (SFR), an initial mass function (IMF) of $dN/dm\propto m^{-2.3}$ \citep{kroupa01} and a MS lifetime of $t_{\rm MS}=10\Gyr(M/\Msun)^{-3}$, we simulate the population of WDs in the galactic disk where for every forming A-type or earlier star ($M>1.5 \Msun$) we assign a $0.17$ chance to be found in a wide orbit with another (lighter) MS companion with $M>1\Msun$ (independent of the companion's specific mass). We find that a fraction of $\sim 60$ per cent of the lighter companions in massive binaries in which the primary already evolved to a WD will also evolve to a WD implying that \emph{$\sim \mathbf{10}$ per cent of WDs have a wide-orbit WD companion.}

These expectations can be directly compared with the statistics of observed companions to WDs. Out of the $\approx 120$ WDs with $M>0.5\Msun$ and $D<20\pc$ presented in \cite{holberg16} we would expect $\sim10$ WODDs and $\sim 5$ wide orbit systems with a WD and a MS companion with $M>1\Msun$. In the observed sample there are $3$ WODDs (WD-0727+482, WD-2226-754 and WD-0747+073) and $4$ WD-MS($M>1\Msun$) systems (Sirius B, Procyon B, WD-1544-377\footnote{The companion of WD-1544-377, HD 140901, mass's uncertainty includes $1\Msun$ \citep{pinheiro14}.} and WD-0415-594). The small number of detected WODDs compared to the expectation is unlikely to be due to a statistical fluctuation. This strengthen's previous suspicions that there are missing WDs in multiple systems in the local sample \citep{ferrario12}. \emph{We expect that about $\sim 10$ WODDs be detected within $20\pc$ in the future.} In particular, the GAIA astrometric mission should eventually detect most of these missing systems by resolving the systems with large separations $P\gtrsim 10\yr$ and finding astrometric solutions for the systems with close separations $P\lesssim 10\yr$. If the fraction of WODDs is established to be much smaller than $10$ per cent, this would raise the interesting possibility that many wide massive binaries become unbound before they become WDs.

Our estimate of the WODD fraction places a tight constraint on the feasibility of the collision model as a primary channel for type Ia supernovae. Following the same assumptions made above and assuming delay time distribution of type Ia SNe of \cite{maoz12, graur13} we find that $\sim 10$ per cent of WODDs should end up with a collision of the WDs in order to account for the SNe rate. This result can be equivalently obtained by assuming production of $0.1$ WDs and $0.001$ type Ia SNe per $\Msun$ of star formation combined with our result that $\sim 10$ per cent of WDs end up in WODDs. This is in tension with the estimate that only a few percent of triple systems with WODDs having the right hierarchy \citep{katz12} lead to a collision. This suggests that in order for the collision model to work, either the majority of WODDs have a relevant tertiary (leaving a modest discrepancy of order $2$) or that other effects such as higher multiplicity \citep{pejcha13} or passing stars \citep{antognini16} substantially increase the chance for collisions in some of the systems.

\section*{Acknowledgements}
We thank Subo Dong, Eran Ofek and Doron Kushnir for useful discussions. This research was supported by the I-CORE Program (1829/12) and the Beracha Foundation.

\bibliographystyle{mnras}
\bibliography{bib_for_massive_binaries_paper}

\bsp
\label{lastpage}
\end{document}